%% file: article.tex
\documentclass[twocolumn]{layout/tudelft-aiaa}

\usepackage[T1]{fontenc}
\usepackage[utf8]{inputenc}

\usepackage{graphicx} 
\usepackage{amsmath} 
\usepackage{siunitx} 
\usepackage{tabularx} 
\usepackage{subcaption} 

\title{Double spending prevention of digital Euros using a web-of-trust}

\author{Atanas Marinov, Jurriaan Den Toonder, Joep de Jong, Pieter Tolsma,\\ Nils van den Honert, Johan Pouwelse (course supervisor)}
\affil{Distributed Systems, Delft University of Technology}
\begin{document}

\AlwaysPagewidth{

\maketitle
\vspace{5pt}
\begin{center}
    — student project —
\end{center}
\vspace{5pt}

\begin{abstract}
    \noindent In order to provide more security on double-spending, we have implemented a system allowing for a web-of-trust. In this paper, we explore different approaches taken against double-spending and implement our own version to avoid this within TrustChain\cite{trust} as part of the ecosystem of EuroToken\cite{eurotoken}, the digital version of the euro. We have used the EVA protocol\cite{eva} as a means to transfer data between users, building on the existing functionality of transferring money between users. This allows the sender of EuroTokens to leave recommendations of users based on their previous interactions with other users. This dissemination of trust through the network allows users to make more trustworthy decisions. Although this provides an upgrade in terms of usability, the mathematical details of our implementation can be explored further in other research.  
\end{abstract}
\vspace{15pt}
}

\input{introduction}
\input{research}
\input{requirements}
\input{protocol}
\input{results-discussion}
\input{conclusion}


\bibliography{article}


\end{document}

%% file: introduction.tex
\section{Introduction}
Double spending is one of the severe problems that the creators of digital currencies attempt to solve. Double spending can be summarized as the endeavor of an individual to exploit the digital payment mechanism to enable himself to spend the same digital token twice with two different distinct users, leaving one of them empty-handed. While it is not hard to prevent it in an online environment, solving it in an offline setting proves to be more complicated or even impossible. Unfortunately, the offline context is where digital currencies find most of their potential uses.

The presence of such a major flaw can significantly undermine the trust in a financial system and cause further inflation, which will devalue the underlying currency. But why do we need digital currency then? Highly-developed societies such as the European Union have been moving in a cashless direction. For example, according to Sweden’s central bank, the Riksbank \cite{ sveriges-riksbank-2020}, the number of people using cash has fallen from 39\% to only 9\% between 2010 and 2020. In a case of a crisis such as large-scale power or internet outage caused, for instance, by a natural disaster that lasts for a long time, all these people will be unable to spend their money.

Instead of hoping that the world will continue to function as normal independent of the factors we cannot control, we decided to focus on the double-spending problem. This research proposes a method, working in offline scenarios, which tackles the double-spending problem by developing a web-of-trust in the network through continuous dissemination of the trustworthy users.

The document is structured as follows: First, research specifying current methods to prevent and detect double-spending are discussed. Then the requirements for the offline euro and the assumptions about our implementation are introduced. Following that, the actual protocol is described in more detail. In the end, a short discussion about the potential consequences of such a system is presented.

%% file: research.tex
\section{Related Work}

Several methodologies have been developed in order to combat double-spending. We have distinguished between two major approaches: double spending prevention and double spending detection. Within these distinctions, we have opted to differentiate between online and offline systems. This way, we can explore disaster-proof solutions already proposed within this problem space.

\subsection{Prevention}
\subsubsection{Online}
There are different types of proposals to deal with the prevention of the double-spending problem in an online setting. The first one is to use blockchain models such as Bitcoin and Ethereum\cite{ethereum}, which use a centralized ledger \cite{Bitcoin}. A strategy to ensure double spending does not occur is to stall the transaction of goods or services until payment is received. If we follow this strategy, we have to wait until our transaction is confirmed in a block that is generally accepted on the chain. The major downside of pursuing such a strategy, which makes it unsuitable for e-cash, is the waiting time associated with the transaction. In everyday transactions, which would compose a majority of all transactions, require payment confirmation within seconds. Delayed payments would significantly hamper commerce. This is generally not feasible with the standalone Bitcoin ledger, where confirming a transaction happens in the order of minutes, rather than seconds. 

One of the first attempts to deal with these was the so-called green addresses. When a sender wants to send money, instead of doing it directly, he contacts his trusted financial intermediary, which subtracts the amount from his account and sends it to the receiving party. Upon receipt, the receiver sees that the money is coming from a trusted intermediary, i.e., a green address, that can be trusted to not double spend. Nonetheless, the enormous amount of trust that was placed on the intermediaries resulted in the concept never truly succeeding. Still, parts of the trust idea could be reused by us. 

A different approach to ensure confirmation broadcasting the transaction to a set of neighbors or a random set of nodes \cite{10.1007/s10207-018-0422-4}. Another set of possible solutions is obtained by using centralization within the system . David Chaum uses this approach in his e-cash scheme to combat double spending \cite{chaum1983blind}. 

\subsubsection{Offline}
The main problem of the schemes previously presented is the online restriction itself, where we always have to be connected to the system in order to spend money. The assumption that a user will always be online is not realistic, considering that a disaster is possible. One solution used in an offline environment is to expose double spenders, creating awareness to others within the network that this person is liable to double spending \cite{chaum1983blind}. This is done by revealing the double spender through the use of cryptography, where there is a high probability that the double spender is caught. This approach to an offline solution has been adopted by many researchers in the field \cite{Cham} \cite{FAN} \cite{kang}. This solution does have drawbacks. It does rely on an trusted instance, which in term creates a supplementary banking and legal system. If decentralization of money is a core principle within our design space, instantiating such a system would harm our ability to fully decentralize the system.  

A different approach to prevent this problem is the use of secure tamper-proof hardware. A device can be constructed in the form of a smart wallet or card, which monitors the balance of the user and updates the balance it if the user receives or spends money. The method has been adopted by many companies, notably MasterCard and Visa. Unfortunately, the technology never succeeded. While the devices itself provided enough security against double spending, the user was still liable to lose their funds in case of theft or by damaging the device. In addition to this, the user can only interact with people or institutions who use and accept the device.

\subsection{Detection}
Detection of double spending can be done in two particular stages of the transaction: before the transaction is performed, known as proactive detection, and after the transaction has happened, which we know as retroactive detection.\par
Proactive detection happens before the transaction is registered, during the chain verification phase. In order to perform this, we need some previously collected of the malicious node and its transaction history. This either happens by having a previous transaction with the malicious node or, more likely, through information dissemination through the network. \par
Retroactive detection happens after the transaction. Via dissemination, we obtain new information regarding sequence numbers, which shows us a double-spending event has occurred.
\subsubsection{Online}

In a centralized way, for example with Bitcoin, we have a lower speed than is usually required in smaller transactions. The usual way of dealing with these fast transactions is accepting the risk of a payment not being approved, which will be a problem for the receiver. It is mitigated by waiting till the payment has been propagated through the network. The receiver monitors a random sample of nodes and waits to see if its payment is occurring in the network, detecting if the payment has actually been accepted. If we take a large enough sample to monitor, we can prevent the majority of double-spending. \cite{cryptoeprint:2012:248} \cite{6688717}

An idea to stop double-spending by malicious nodes in a pair-based ledger is to anonymize requests to view a ledger. This way, the malicious node cannot handcraft a faulty chain, since it does not know what the node knows it is trying to cheat on. The intermediary anonymizer nodes would shield the communication between the two nodes. It still poses the risk that the anonymizer node itself is malicious. In order to guarantee fairness, the system would require anonymizer nodes to be audited. \cite{mohammad2019enabling}

\subsubsection{Offline}
In general, there seems to be limited research done on double-spending detection in an offline environment. One of the systems proposed is an off-line karma system, where tokens need to be reminted after a certain time, in order to validate them for extended offline periods. \cite{garcia2005off}

%% file: requirements.tex
\section{Offline Euro Requirements}
In order to bound our solution space, we need to identify some constraints. Aside from working with TrustChain, we have identified some other constraints that influence our design space. A list can be found below:

\begin{itemize}
    \item Fully offline-capable 
    \item Completely distributed
    \item Permissionless
    \item Pseudo-anonymous
    \item Independent of other authorities - legal, bank
\end{itemize}

Given the research in the previous section and the constraints, we have decided to opt for double-spending detection, as it is most feasible in an offline environment. Double-spending prevention is generally difficult even in an online requirement, as it also requires waiting for dissemination of the transaction through the network. Only then can the purchased goods or services be delivered.

In short, we want to be able to share trustworthiness across the network, allowing us to check if our peers are known for previous good behavior, in essence, known and therefore trusted not to double-spend. If we see others validate them upon the exchange of trust scores, the user is indicated of their reliance within the network. This in turn allows the user to make informed decisions regarding sending or receiving money from this peer. 

To summarize the must-haves:
\begin{itemize}
    \item Allow storage of key-value pairs with public keys of peers
    \item Allow users to share their knowledge of others during interactions
    \item Communicate these scores to the user when interacting with a score with a wallet
    \item Create some form of computation towards a trust score 
\end{itemize}   
And the wont-haves:
\begin{itemize}
    \item Definitive formula on how to calculate trust based on the transaction graph and previously known attacks
    \item Blocking of users from the network
\end{itemize}

%% file: protocol.tex
\section{Offline Peer-to-Peer payment protocol}

\begin{figure}[ht]
  \centering
  \includegraphics[scale=0.23]{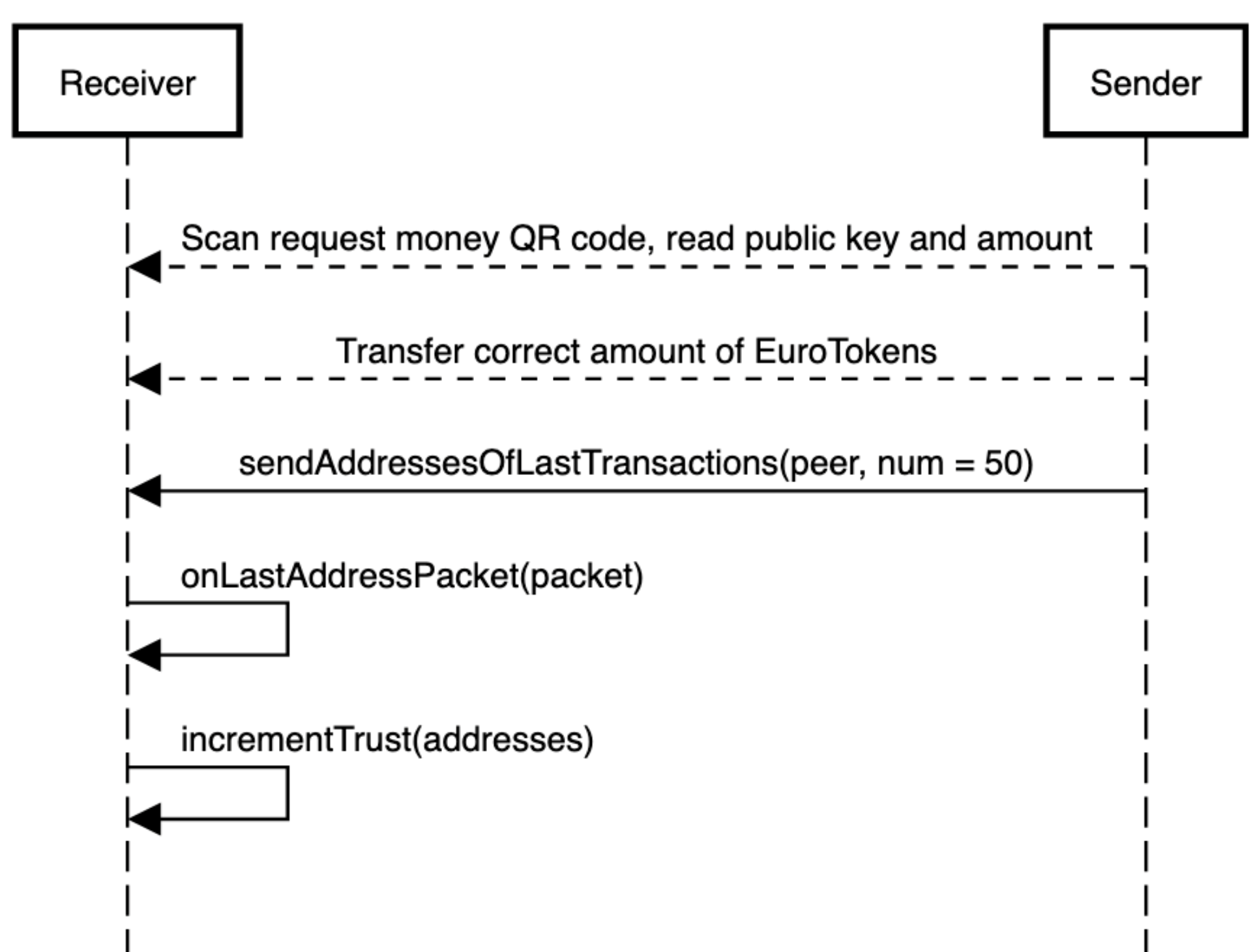}
  \caption{Workflow of the app}
  \label{fig:flow}
\end{figure}

\begin{figure}[ht]
  \centering
  \includegraphics[scale=0.5]{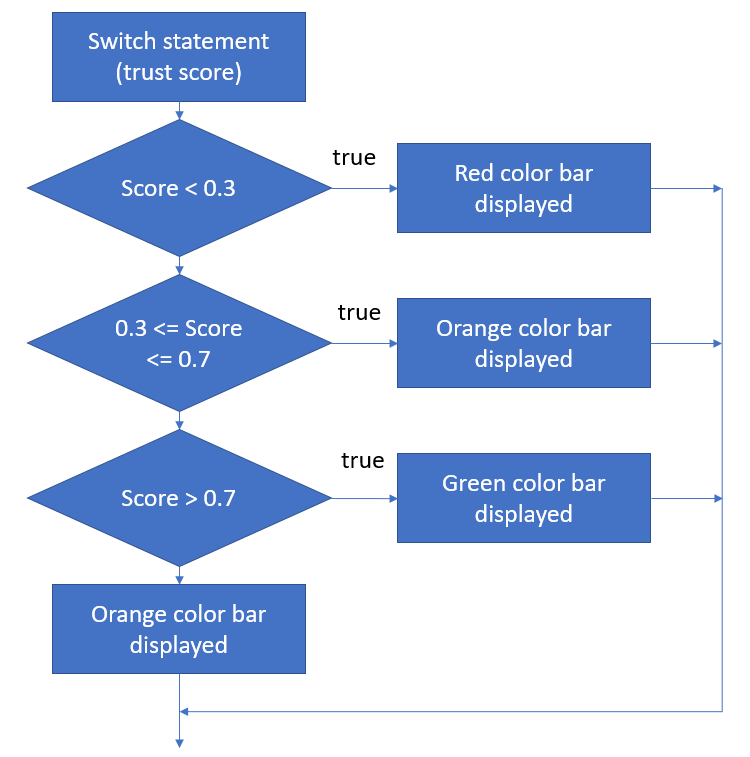}
  \caption{Flow diagram for displaying information regarding trust scores}
  \label{fig:switch}
\end{figure}

\begin{figure}[ht]
    \centering
    \includegraphics[scale=0.7]{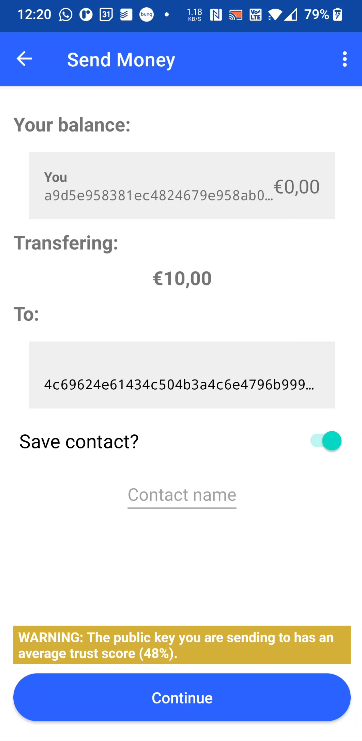}
    \caption{Screen capture of a transaction showing a trust score}
    \label{fig:screenshot}
\end{figure}

The workflow of the EuroToken app is illustrated in Figure \ref{fig:flow}. The first two steps are related to the general money transfer mechanism that was implemented beforehand. Our contributions can be summarised in the last three steps. 
Each user keeps a database consisting of the public keys of wallets, together with a score between 0 and 100. Here, an unknown public key receives a trust score of 0, while 100 symbolizes maximum available trust. When the users want to send money, they scan the QR of the receiving party. Upon trying to complete a transaction, we go through our collected records and see if we can find a matching public key of the peer we are interacting with. If we cannot find this, the user is informed that we currently have no information on this user. Alternatively, if we do find a key pair, we read the trust score from the database. We define some thresholds, as can be seen in Figure \ref{fig:switch}, where we give a color indication based on the value associated with the key. How this would look to the end-user is shown in Figure \ref{fig:screenshot}. 

After interacting with another user, the receiver gets a list of the last 50 users the sender has interacted with. As the legitimacy of the transaction is dependent only on the truthfulness of the sender, no public keys are sent in the other direction. Build on the IPv8 protocol, a P2P networking layer \cite{ipv8documentation}, the EVA protocol is utilized to transfer the list. Sending list has three phases - first, the public keys are serialized and deserialized into a correct payload, then an encrypted packet containing the payload and message identifier is constructed, and finally, the binary is transmitted. Upon receipt, if there are familiar users within this list, we can update their trust scores. For now, this is simply done by incrementing their trust score by one percentage point in the database. If we encounter a new public key within the sent list, we add this person to our database. Over time, this will allow us to see nodes within the network that thoroughly interact with other users, which we intuitively trust more than users with limited interaction.

Our payload typically consists of 50 public keys, each comprised of 148 hexadecimal characters. The means each key is $148*4=592$ bits, making our payload $592*50=29600$ bits. Each transaction would then only be 0.0037 MB, which is therefore not a heavy load on the system in general, even encrypted and packaged over IPv8. 

%% file: results-discussion.tex
\section{Results \& Discussion}
Although the system sets out what was intended by us from a technical viewpoint, it still suffers from some difficult problems. Further research will have to answer these problems. It does however show the technological capability to propagate trust through the system. 

\subsection{Reputation scores}
One issue is that the trust score could lead to a sort of social credit system, similar to Chinese implementations. These could infer social implications, as users might get excluded due to their low trust scores. These social outcasts could theoretically only be accepted by each other, leading to double-spending regions within the network. This is not desired, as they could be excluded from a majority of their desired transactions. Similar consequences can be observed in credit scores in America \cite{doi:10.1007/s12114-015-9215-4} \cite{BUBB20141}. Not meeting a certain threshold can exclude you from lending or general business transactions.

\subsection{Security concerns}
Other issues are more closely related to the security of these trust scores. A high trust score could become a desirable trait since more people would trust the user to do business with them. Once this happens, people will try to game the system. One possible adversary strategy would be to boost your own scores by repeatedly transferring between two wallets (or a similar cycle within a graph), resulting in a high trust score. Similar approaches can be seen in other fields, such as search algorithms \cite{Disinfo}. For now, the algorithms creating such scores are generally hidden and in constant change. This security by obscurity approach could be also applied to the trust score algorithm. By not using a linear function, but for example machine learning, we could obfuscate some of the factors that the algorithm considers predictive for double-spending. If these factors remain unknown, bad actors have more troubles understanding and manipulating the system versus a publicly known algorithm. A major downside of security by obscurity is that once the method of calculating trust scores is known, it can be easily taken advantage of to boost your own score.

%% file: conclusion.tex
\section{Conclusion}

As we have seen, it is possible to create an additional layer of trust to improve security against double-spending. This setup allows us to create a web-of-trust, where individual users can recommend interacting with other users. The technical aspects of our implementation still could use some improvements. This can also help new users of the system manage their risk tolerance and overall understanding of the decentralized ledger system. This in turn improves overall adoption. \newline 
Some further research needs to be done on how to calculate this score, beyond our linear additive approach. We could opt for more sophisticated methods, which allow less misuse of the system and faster dissemination of bad actors within the system.  Another action that needs to be implemented is tying an update to connect a score update to the corresponding transaction. This prohibits adversaries from continuously updating scores. Additional research could also be done into the prevention of cycle boosting trust scores.